\documentclass{elsarticle}

\usepackage{natbib}
\usepackage{amssymb}
\usepackage{graphicx}
\usepackage{color}
\journal{Physica A}

\begin{document}

\begin{frontmatter}

\title{{\bf Including transcription factor information in the superparamagnetic clustering of microarray data}}

\author[label1]{M. P. Monsiv\'ais-Alonso}\ead{monsivais@ipicyt.edu.mx}
\author[label1]{$\ $J. C. Navarro-Mu\~noz}\ead{jcarlos@ipicyt.edu.mx}
\author[label2]{L. Riego-Ruiz}\ead{lina@ipicyt.edu.mx}

\author[label1]{R. L\'opez-Sandoval}\ead{sandov@ipicyt.edu.mx} 
\author[label1]{H.C. Rosu}\ead{hcr@ipicyt.edu.mx}

\address[label1]{Division of Advanced Materials, IPICyT, Instituto Potosino de Investigacion Cientifica y Tecnologica, San Luis Potos\'{\i}, S.L.P., Mexico}
\address[label2]{Division of Molecular Biology, IPICyT, Instituto Potosino de Investigacion Cientifica y Tecnologica, San Luis Potos\'{\i}, S.L.P., Mexico }

\begin{abstract}
In this work, we modify the superparamagnetic clustering algorithm (SPC) by adding an extra weight to the interaction formula that considers which genes are regulated by the same transcription factor. With this modified algorithm that we call SPCTF, we analyze Spellman {\it et al.} microarray data for cell cycle genes in yeast, and find clusters with a higher number of elements compared with those obtained with the SPC algorithm. Some of the incorporated genes by using SPCFT were not detected at first by Spellman {\it et al.} but were later identified by other studies, whereas several genes still remain unclassified. The clusters composed by unidentified genes were analyzed with MUSA, the motif finding using an unsupervised approach algorithm, and this allow us to select the clusters whose elements contain cell cycle transcription factor binding sites as clusters worth of further experimental studies because they would probably lead to new cell cycle genes. Finally, our idea of introducing available information about transcription factors to optimize the gene classification could be implemented for other distance-based clustering algorithms.

\end{abstract}

\begin{keyword}
Superparamagnetic clustering \sep similarity measure \sep microarrays \sep cell cycle genes \sep transcription factors.\\

\bigskip
\bigskip
paper-physa-3 20100901.tex \hfill Physica A 389(24), 5689-5697 (2010)\\
\hfill doi: 10.1016/j.physa.2010.09.006 
\end{keyword}

\end{frontmatter}

\newpage

\section{Introduction}
DNA microarrays allow the comparison of the expression levels of all genes in an organism in a single experiment, which often involve different conditions ({\it i.e.} health-illness, normal-stress), or different discrete time points ({\it i.e.} cell cycle) \cite{Schena 1995, Schulze 2001}. Among other applications, they provide clues about how genes interact with each other, which genes are part of the same metabolic pathway or which could be the possible role for those genes without a previously assigned function. DNA microarrays also have been used to obtain accurate disease classifications at the molecular level \cite{McLachlan 2004, Trevino 2007, Draghici 2003}. However, transforming the huge amount of data produced by microarrays into useful knowledge has proven to be a difficult key step \cite{Fadiel 2003}.

On the other hand, clustering techniques have several applications, ranging from bioinformatics to economy \cite{cluster_app1, cluster_app2, cluster_app3}. Particularly, data clustering is probably the most popular unsupervised technique for analyzing microarray data sets as a first approach. Many algorithms have been proposed, hierarchical clustering, k-means and self-organizing maps being the most known \cite{clustering_rev1, clustering_rev2}. Clustering consists of grouping items together based on a similarity measure in such a way that elements in a group must be more similar between them than between elements belonging to different groups. The similarity measure definition, which quantifies the affinity between pairs of elements, introduces {\it a priori} information that determines the clustering solution. Therefore, this similarity measure could be optimized taking into account additional data acquired, for example, from real experiments. Some works with {\it a priori} inclusion of bioinformation in clustering models can be found in \cite{Pan, HuangPan}.

In the case of gene expression clustering, the behavior of the genes reported by microarray experiments is represented as $N$ points in a $D$-dimensional space, being $N$ the total number of genes, and $D$ the number of conditions. Each gene behavior (or point) is then described by its coordinates (its expression value for each condition). Genes whose expression pattern is similar will appear closer in the $D$-space, a characteristic that is used to classify data in groups. In our case, we have used the Superparamagnetic Clustering Algorithm (SPC) \cite{Blatt, Blatt97, Wiseman,Domany}, which was proposed in 1996 by Domany and collaborators as a new approach for grouping data sets. However, this methodology has difficulties dealing with different density clusters, and in order to ameliorate this, we report here some modifications of the original algorithm that improve cluster detection. Our main contribution consists on increasing the similarity measure between genes by taking advantage of transcription factors, special proteins involved in the regulation of gene expression.

The present paper is organized as follows: in Section 2, the SPC algorithm is introduced, as well as our proposal to include further biological information and our considerations for the selection of the most natural clusters. Results for a real data set, as well as performance comparisons, are presented in Section 3. Finally, Section 4 is dedicated to a summary of our results and conclusions.

\section{Method}

\subsection {Superparamagnetic Clustering Algorithm (SPC)}

A Potts model can be used to simulate the collective behavior of a set of interacting sites using a statistical mechanics formalism. In the more general inhomogeneous Potts model, the sites are placed on an irregular lattice. Next, in the SPC idea of Domany {\it et al.} \cite{Blatt}, each gene's expression pattern is represented as a site in an inhomogeneus Potts model, whose coordinates are given by the microarray expression values. In this way, a particular lattice arrangement is spanned for the entire data set being analyzed.

A spin value $\sigma_i$, arbitrarily chosen from $q$ possibilities, is assigned to each site, where $i$ corresponds to the site of the lattice $i=1,2,..., N$. The main idea is to characterize the resulting spin configuration by the ferromagnetic Hamiltonian:
\begin{equation}
\label{eq:Ham}
H=-\sum_{i,j} J_{ij}\delta_{\sigma_i,\sigma_j}, \qquad \sigma_i=1,...,q,
\end{equation}
where the sum goes over all neighboring pairs, $\sigma_i$ and $\sigma_j$ are spin values of site $i$ and site $j$ respectively, and $J_{ij}$ is their ferromagnetic interaction strength.

Each site interacts only with its neighbors, however since the lattice is irregular, it is necessary to assign the set of nearest-neighbors of each site using the so-called $k$-mutual-nearest-neighbor criterion \cite{kmNN}. The original interaction strength is as follows:
\begin{equation}\label{eq:js}
J_{ij}= \left\{ \begin{array}{ll}
\frac{1}{\hat K}e^{-\frac{d_{ij}^2}{2a^2}} & \textrm{if $i$ and $j$ are neighbors}\\
\\
0 & \textrm{otherwise,}
\end{array} \right.
\end{equation}
with $\hat K$ the average number of neighbors per site and $a$ the average distance between neighbors. The interaction strength between two neighboring sites decreases in a Gaussian way with distance $d_{ij}$ and therefore, sites that are separated by a small distance have more probability of sharing the same spin value during the simulation than the distant sites. On the other hand, said probability, $P_{ij} = (1 - e^{(-J_{ij}/T)})$, also depends on the temperature $T$, which acts as a control parameter. At low temperatures, the sites tend to have the same spin values, forming a ferromagnetic system. This configuration is preferred over others because it minimizes the total energy. However, the probability of encountering aligned spins diminishes as temperature increases, and the system could experience either a single transition to a totally disordered state (paramagnetic phase), or pass through an intermediate phase in which the system is partially ordered, which is known as the superparamagnetic phase. In the latter case, varios regions of sites sharing the same spin value emerge. Sites within these regions interact among them with a stronger force, exhibiting at the same time weak interactions with sites outside the region. These regions could fragment into smaller grains, leading to a chain of transitions within the superparamagnetic phase until the temperature is so high that the system enters the paramagnetic phase, where each spin behaves independently. This hierarchical subdivision in magnetic grains reflects the organization of data into categories and subcategories. Regions of aligned spins emerging during simulation correspond to groups of points with similar coordinates, {\it i.e.}, similar gene expression patterns \cite{Blatt, Blatt97, Wiseman}. This subdivision can be simulated, for example, by using the Monte Carlo approach, by which one can compute and follow the evolution of system properties such as energy, magnetization and susceptibility, while the temperature is modified. In addition, the temperature ranges in which each phase transition takes place can be localized.

Rather than thresholding the distances between pairs of sites to decide their assignment to clusters, the pair correlation $G_{ij}$, indicating a collective aspect of the data distribution, is preferred . It can be calculated as follows \cite{Blatt97}
\begin{equation}
\label{eq:corre}
G_{ij} = \frac{(q-1)(\langle \delta_{\sigma_i,\sigma_j} \rangle)+1}{q-1} ~.
\end{equation}

In this way, $G_{ij}$ is the normalized probability for f\mbox{}inding two Potts spins $\sigma_i$ and $\sigma_j$ sharing the same value for a given temperature step. If both spins belong to the same ordered region, their correlation value would be close to one, otherwise their correlation would be close to zero \cite{Domany}. Thus, for each temperature step, two sites are assigned to the same cluster if their correlation exceeds a threshold value of $G_{ij}>0.5$. If a site does not have a single correlation value greater than $0.5$, it is joined with its neighbor showing the highest value.\\

\subsection{Transcription Factors in SPC (SPCTF)}

For our SPCTF algorithm, we also accept sites whose $G_{ij}$ are larger than $0.5$ in order to build a cluster. However, differently from the traditional SPC algorithm \cite{Blatt, Blatt97, Wiseman, Domany}, if two sites do not reach the $G_{ij}$ value greater than $0.5$ they are not connected. This is because with our data we have found that the original condition led to unnatural growth of some clusters when the temperature is increased.

As already mentioned, the data are fragmented in various clusters for each temperature value, and for higher temperatures, the number of clusters increases due to finer and finer segmentation. In order to select the more representative clusters through all temperature steps, we assign a stability value to each obtained cluster, based on its evolution. We define $T_{t}$ as the number of temperature steps until the system reaches the paramagnetic phase and $T_{v}$ as the number of temperature steps a cluster $v$ survives, while $I_t$ and $I_{v}$ are defined as the total number of sites and the number of elements in a given cluster, respectively. We assign a stability parameter ${\cal S}_{v}$ to each cluster, as follows:

\begin{eqnarray}
\label{eq:stability}
{\cal S}_{v}&=& \frac{ col_{v}row_{v} }{ \mid col_{v}-row_{v} \vert  + \epsilon }, \quad
\end{eqnarray}
where $col_{v}=\frac {T_{v}}{T_{t}}$ is the fraction of temperature steps a cluster $v$ survives, while $row_{v}=\frac {I_{v}}{I_{t}}$ is the fraction of total elements belonging to $v$. The advantage of using the stability parameter ${\cal S}_{v}$ is that it gives preference to clusters that survive several temperatures, but also have an acceptable number of elements. We added a small positive real number $\epsilon$ to the denominator in the expression of ${\cal S}_{v}$ for the special case when $col_v=row_v=n$, where $n$ belongs to the range $(0, 1]$, leading to ${\cal S}_{v}=\frac{n^2}{\epsilon}$ instead of the infinity.

It has been reported that the main drawback of the SPC algorithm consists of dealing with data showing regions of different density \cite{Ott, Stramaglia}. In this case, either depending on temperature or the number of neighbors selected, some clusters will easily get prominent whereas the detection of others will be hindered. To overcome this problem, at least two techniques have been proposed {\it e.g.}, sequential superparamagnetic clustering \cite{Ott} and a modularity approach \cite{Stramaglia}. Our idea is to take advantage of already available biological information to improve lattice connectivity in such a way that biologically significant clusters have more probability of being detected by the algorithm.

Indeed, at the transcriptional level, the expression of a gene could be promoted/suppressed by the binding of the proteins named transcription factors to specific sequences on the gene promoter region. Then, if a group of genes shows the same expression behavior in a microarray experiment, it is quite possible that they are being regulated by a specific transcription factor, forming a group of coregulated genes \cite{Yu}. Thus, available information about which genes are targeted by the same transcription factors may be useful in the detection of groups of genes with similar expression profiles.

To make effective this idea, we downloaded from {\it www.yeastract.com} a list of yeast transcription factors that are well documented, and whenever two neighboring genes are controlled by the same transcription factor, we increased their interaction strength. It is important to note that the list provided by {\it www.yeastract.com} includes transcription factors associated with several processes and are not only cell cycle related. The formula that takes this into account replaces Eq.~(\ref{eq:js}) of the original algorithm, and has the following form:
\begin{equation}\label{eq:jsmodified}
J_{ij}= \left\{ \begin{array}{ll}
\frac{F}{\hat K}e^{-\frac{d_{ij}^2}{2(Fa)^2}} & \textrm{if $i$ and $j$ are neighbors},\\
\\
0 & \textrm{otherwise.}
\end{array} \right.
\end{equation}
Here, $F=fn$ is the number of common transcription factors shared by $i$ and $j$ ($n$, which varies for each pair of neighboring genes), multiplied by a factor $f$ which was chosen to be 2.0 after comparing the results obtained with several other values. The selected value has the characteristic of preserving well-defined susceptibility peaks as well as obtaining larger clusters. The objective is to strengthen some connections without preventing the natural fragmentation of clusters caused by the temperature parameter. If two elements do not share a transcription factor, then $F = 1$, recovering the original SPC formula. Therefore, the modified interaction strength between each site and its neighbors is governed by two aspects: the distance between them, which comes from gene expression values generated through microarray experiments, and the number of transcription factors regulating both genes, obtained from documented biological data. Any time two genes share a transcription factor, their interaction strength becomes larger, and this favors that the clusters including these sites remain stable for longer temperature ranges, with the corresponding increase of their stability values.

\section{Results and Discussion}

We analyzed Spellman {\it et al.} \cite{Spellman} microarray data in which gene expression values from synchronized yeast cultures were obtained at various time moments, aiming to identify cell cycle genes. Yeast cultures were synchronized by three methods: adding alpha pheromone, which arrests cells in the G1 phase; using centrifugal elutration for separating small G1 cells; and using a mutation that arrests cells late in mitosis at a given temperature. Combining the three experiments and using Fourier and correlation algorithms, Spellman {\it et al.} \cite{Spellman} reported $800$ cell cycle regulated genes.

The goal was to compare the performance of SPC and SPC with transcription factors (SPCTF), which are algorithms that do not make assumptions about periodicity. Nonetheless, the overall analysis is time consuming and we only selected the data set treated with the alpha pheromone, available at {\it http://cellcycle-www.stanford.edu}. Genes with missing values were discarded, leaving an input matrix of $4489$ genes and $18$ time courses that included only $613$ of the genes reported by Spellman {\it et al.} \cite{Spellman}. Furthermore, as we do not include the other two synchronization experiments, we expect to loose some of their cell cycle genes.

It is worth mentioning that Getz {\it et al.} \cite{Getz} also analyzed the Spellman alpha synchronized set with the SPC algorithm. They took $2467$ genes which have characterized functions and introduced a Fourier transform to take into account the oscillatory nature of the cell cycle. In our case, however, we decided not to introduce any considerations about the periodicity of the data, mainly because the time series cover only two cell cycle periods \cite{Amato}.

We obtain compact gene clusters implementing SPC original algorithm and SPCTF, both with parameter values $k=8$ and $q=20$. The cluster with the highest stability value contains an extremely large number of elements without a clear biological linkage between them. It is mainly composed of genes whose expression do not change significantly over time, thus it is possible that they are included here for this very reason. We discard this cluster from our analysis, although it could always be taken apart and analyzed again with SPCTF by choosing the appropiate number of neighbors to obtain more information.

To compare in more detail both approaches, it is necessary to correlate each cluster in the SPC method with its equivalent in SPCTF. In order to do this, we calculate the euclidian distance between the mean position vector of every cluster in each approach, and choose the pairs with the shortest distance between them. (We recall that the mean position vector of a cluster is obtained by averaging each coordinate between all its elements). Although different measures could have been used, this one performed adequately, as can be seen in the supplementary information file, where we provide a more detailed comparison between SPCTF and SPC clusters. In Table \ref{tabla1}, we present the differences in cluster size as well as the hits, the number of genes reported by Spellman {\it et al.} \cite{Spellman}, which have been included in the clusters. When going through the SPCTF approach, one can see that the first largest cluster looses some genes, while the number of the rest of the clusters augments. Besides, hits or coincidences with Spellman {\it et al.} \cite{Spellman} cell cycle genes in clusters of six or more elements increase by $61\%$, from $108$ to $174$. Therefore, we were able to incorporate several genes to these clusters, mainly from outliers.

\begin{table}[h]
{\bf Comparison between SPC and SPCTF}
\begin{center}
\resizebox{\textwidth}{!}{
\begin{tabular}{|c|c|c|c|c|c|c|c|c|c|c|}
\hline
\hline
 Method  & First & Cluster & Cluster & Cluster & Cluster & Cluster &  Cluster & Total & Total & \bf{Total}  \\
         & Cluster & size $\geq $ 6 & size $=$ 5 & size $=$ 4 & size $=$ 3 & size $=$ 2 & size $=$ 1  & Clusters  & Genes & \bf{Hits} \\
\hline
SPC & 1(2078) \bf{68} & 19(220) \bf{108} & 5(25) \bf{2} & 23(92) \bf{11} & 57(171) \bf{39} & 144(288) \bf{49} & 1615 \bf{336} & 1864 & (4489)  & \bf{613} \\
SPCTF & 1(1657) \bf{64} & 27(359) \bf{174} & 13(65) \bf{23} & 32(128) \bf{22} & 61(183) \bf{30} & 187(374) \bf{60} & 1723 \bf{240} & 2044 & (4489)& \bf{613} \\
\hline
\hline
\end{tabular} }
\end{center}
\caption{Number of clusters for different cluster size. The total number of genes for each cluster size appears in parentheses and their hits with Spellman {\it et al.} \cite{Spellman} appear in bold type. Hits with the $613$ cell cycle genes reported by Spellman {\it et al.} \cite{Spellman} increase for clusters of size $6$ and bigger, while decreasing in the first cluster and outliers.}
\label{tabla1}
\end{table}

\begin{figure}[!thp]
\begin{center}
\includegraphics[height=56mm]{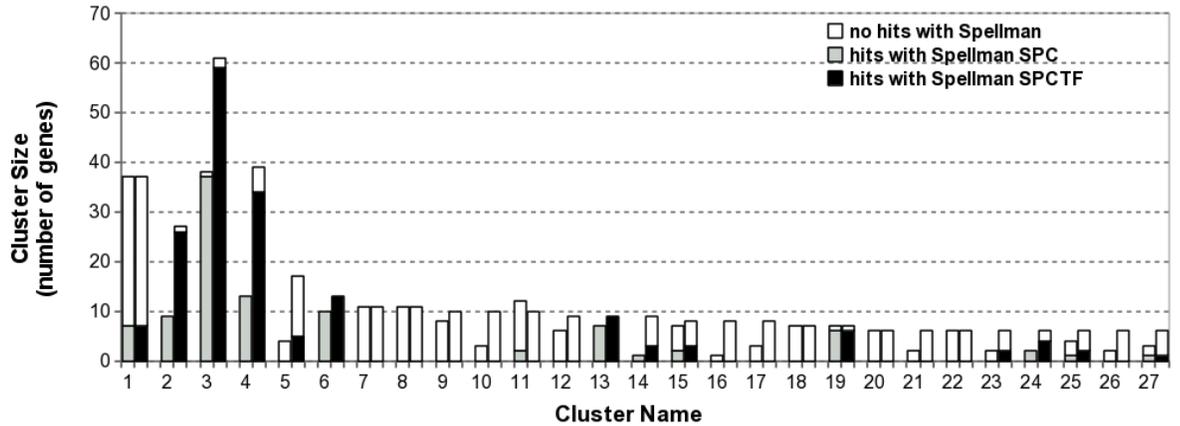}
\caption{General comparison of the first $27$ clusters, discarding the first one. Gray bars correspond to the clusters obtained with the SPC algorithm and black bars to the equivalent clusters in SPCTF. Groups tend to increase in size and also in hits with cell cycle genes reported by Spellman {\it et al.} \cite{Spellman}, with the exception of cluster $11$.}
\label{fig:todos_cc}
\end{center}
\end{figure}

In the following analysis, we focus on clusters of six or more elements, because we are interested in finding groups of several genes sharing the same expression pattern (coregulated genes). Results of the comparison for the first $27$ most stable clusters, discarding the first one, are shown in Fig.~\ref{fig:todos_cc}. Generally, these clusters incorporate more elements with SPCTF, including more cell cycle genes as those reported by Spellman {\it et al.} \cite{Spellman} and thus improving the matching.

\begin{figure}[!htp]
\begin{center}
\includegraphics[height=56mm]{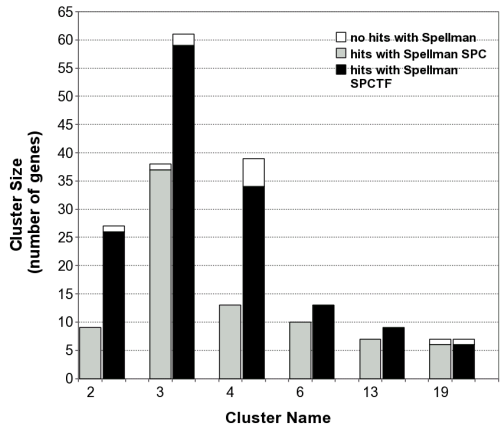}
\caption{Comparison between the SPC and SPCTF results, showing the CC clusters. Gray bars correspond to the clusters obtained with the SPC algorithm and black bars to the equivalent clusters in SPCTF.}
\label{fig:cc_sp}
\end{center}
\end{figure}

\begin{figure}[!htp]
\begin{minipage}[t]{7.0cm}
\includegraphics[height=56mm]{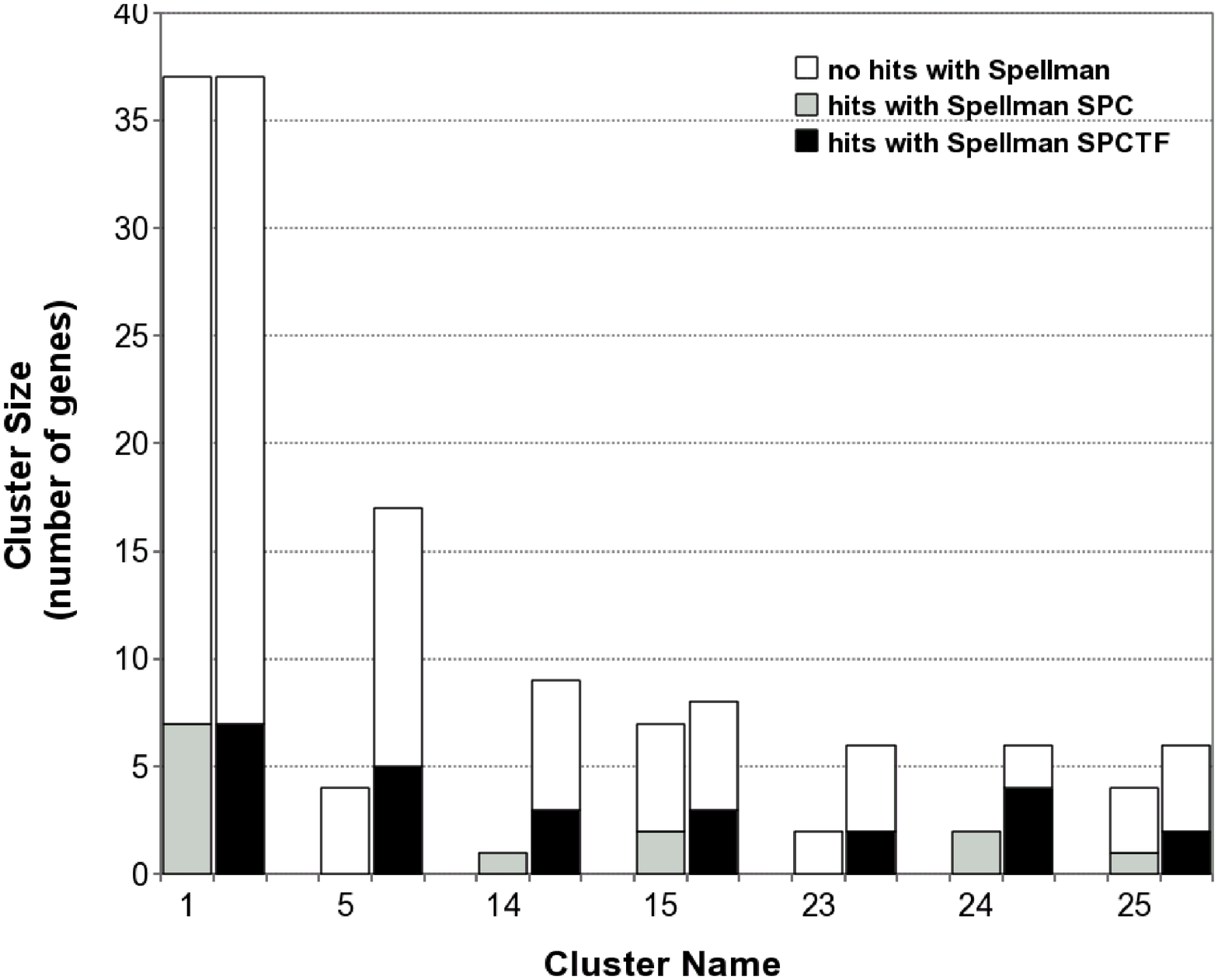}
\end{minipage}
\hfill
\begin{minipage}[t]{9.0cm}
\includegraphics[height=56mm]{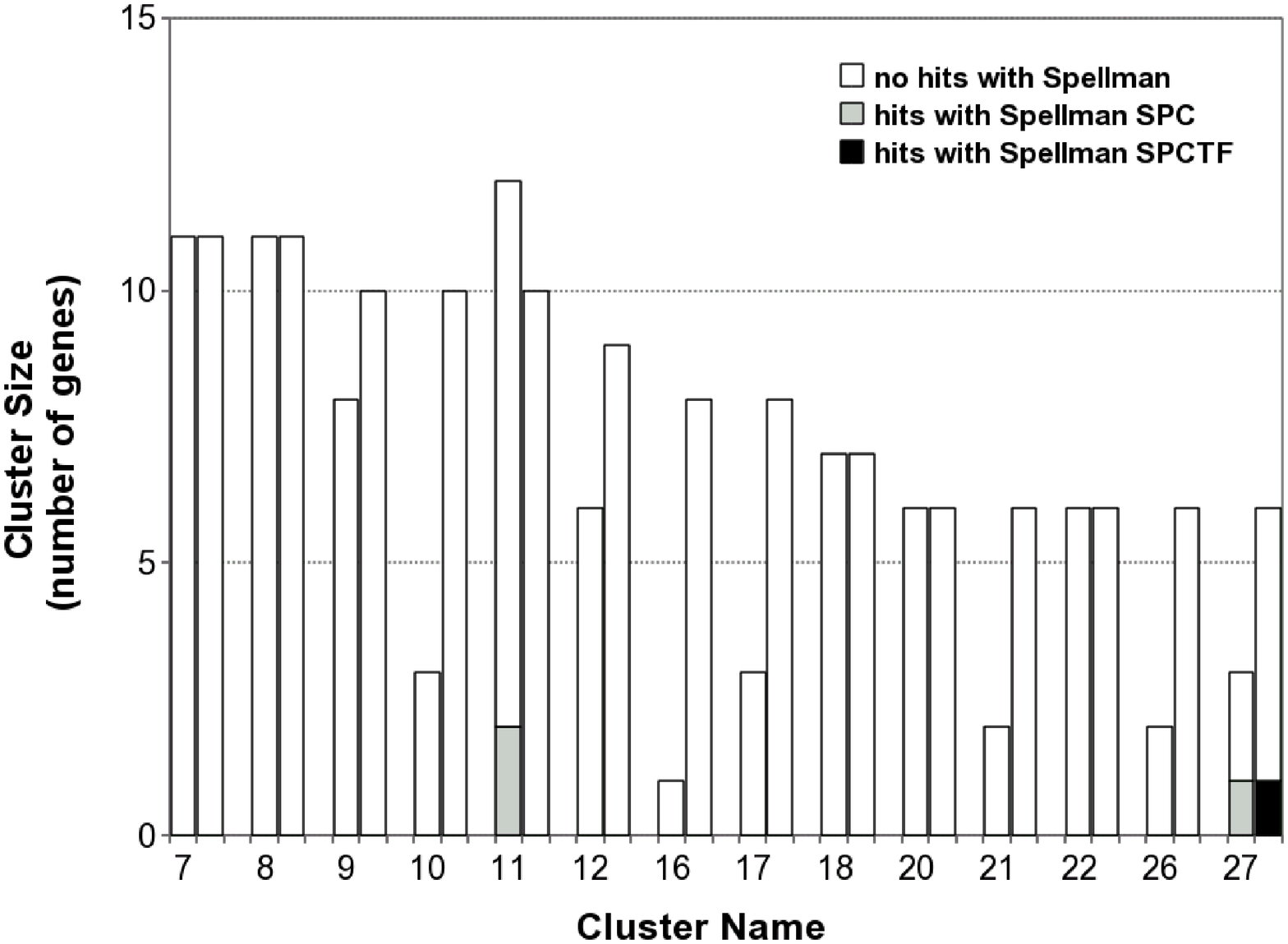}
\end{minipage}
\caption{M and N clusters, left and right respectively. Gray bars correspond to the clusters obtained with the SPC algorithm and black bars to the equivalent clusters in SPCTF.}
\label{fig:m_n}
\end{figure}

Depending on the available information about the genes, we classify the clusters in three groups. The first cluster type, cell cycle genes, CC, corresponds to groups formed in their majority ($\geq 85\%$) by already reported cell cycle genes (Fig.~\ref{fig:cc_sp}). The second type, mixed genes, M, contains clusters with non-reported genes as well as already known cell cycle genes (Fig.~\ref{fig:m_n}), and in the third type, no hits, N, we include the clusters that contain only one hit or are entirely composed of non-previously identified cell cycle genes (Fig.~\ref{fig:m_n}).

\begin{figure}[!thp]
\begin{center}
\includegraphics[height=56mm]{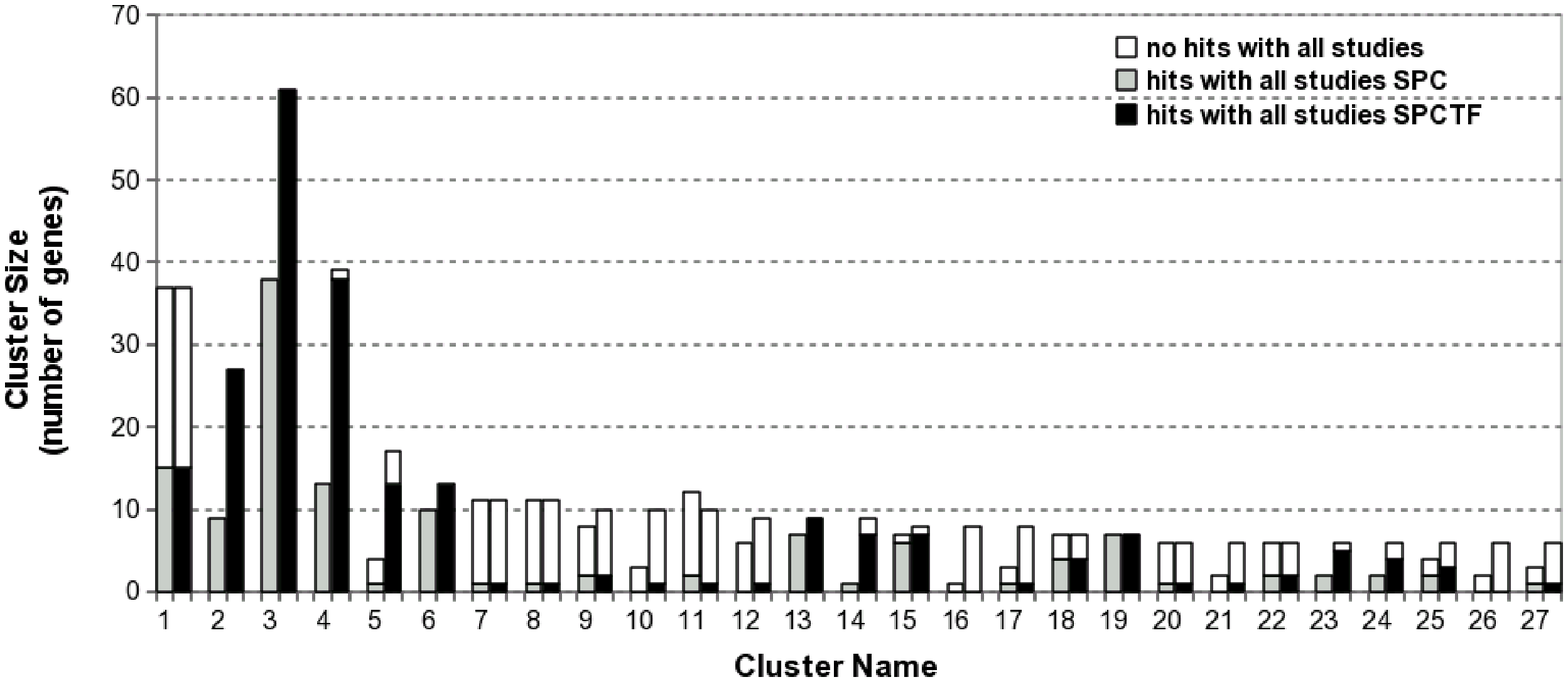}
\caption{General comparison of the first $27$ most stable clusters. Hits are now taken as cell cycle genes reported by all studies. Gray bars correspond to the clusters obtained with the SPC algorithm and black bars to the equivalent clusters in SPCTF.}
\label{fig:all_studies1}
\end{center}
\end{figure}

\begin{figure}[!htp]
\begin{center}
\includegraphics[height=56mm]{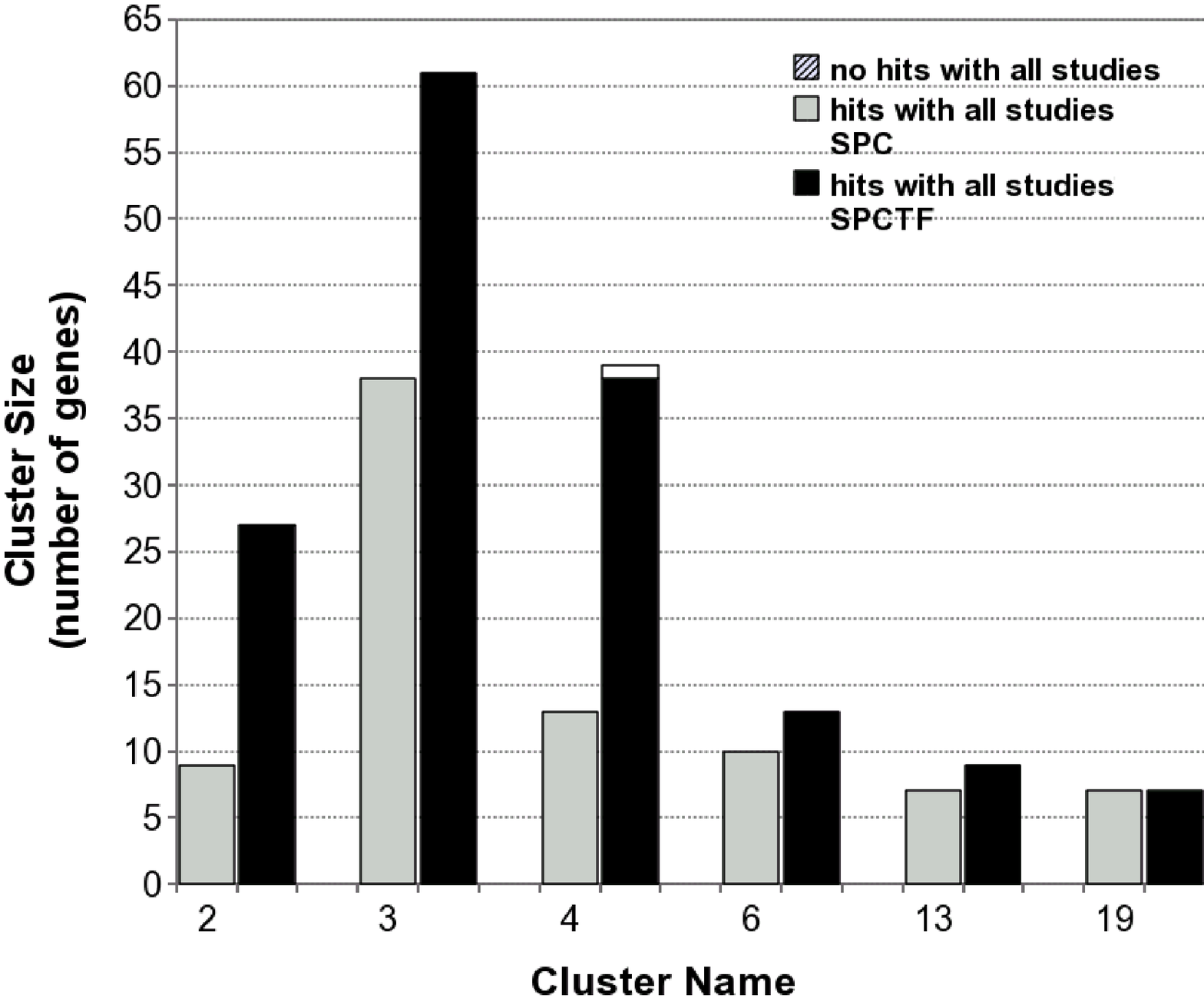}
\caption{Comparison between SPC and SPCTF results, showing CC clusters. Gray bars correspond to the clusters obtained with the SPC algorithm and black bars to the equivalent clusters in SPCTF.}
\label{fig:cc_all}
\end{center}
\end{figure}

\begin{figure}[!htp]
\begin{minipage}[t]{7.0cm}
\includegraphics[height=56mm]{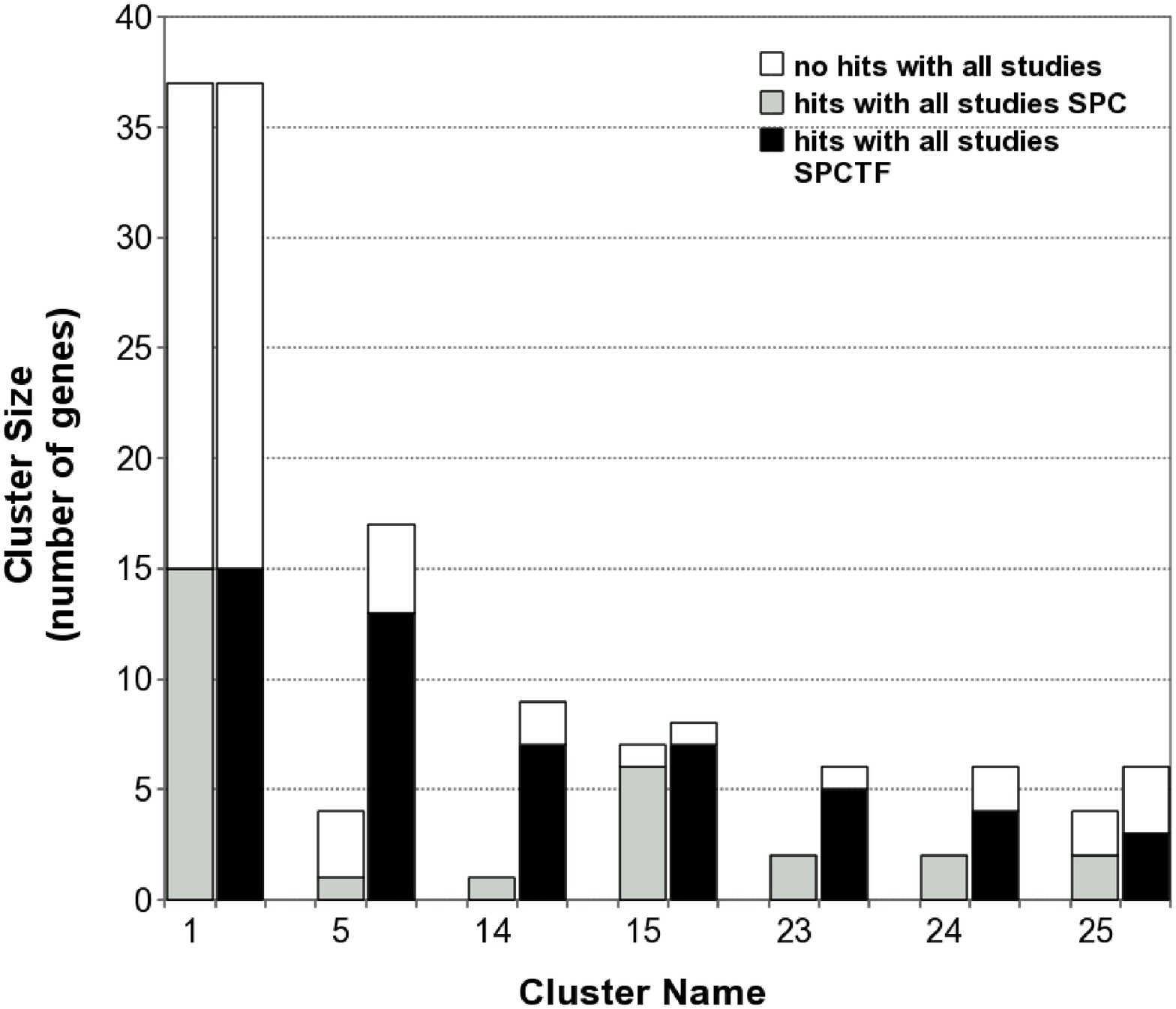}
\end{minipage}
\hfill
\begin{minipage}[t]{9.0cm}
\includegraphics[height=56mm]{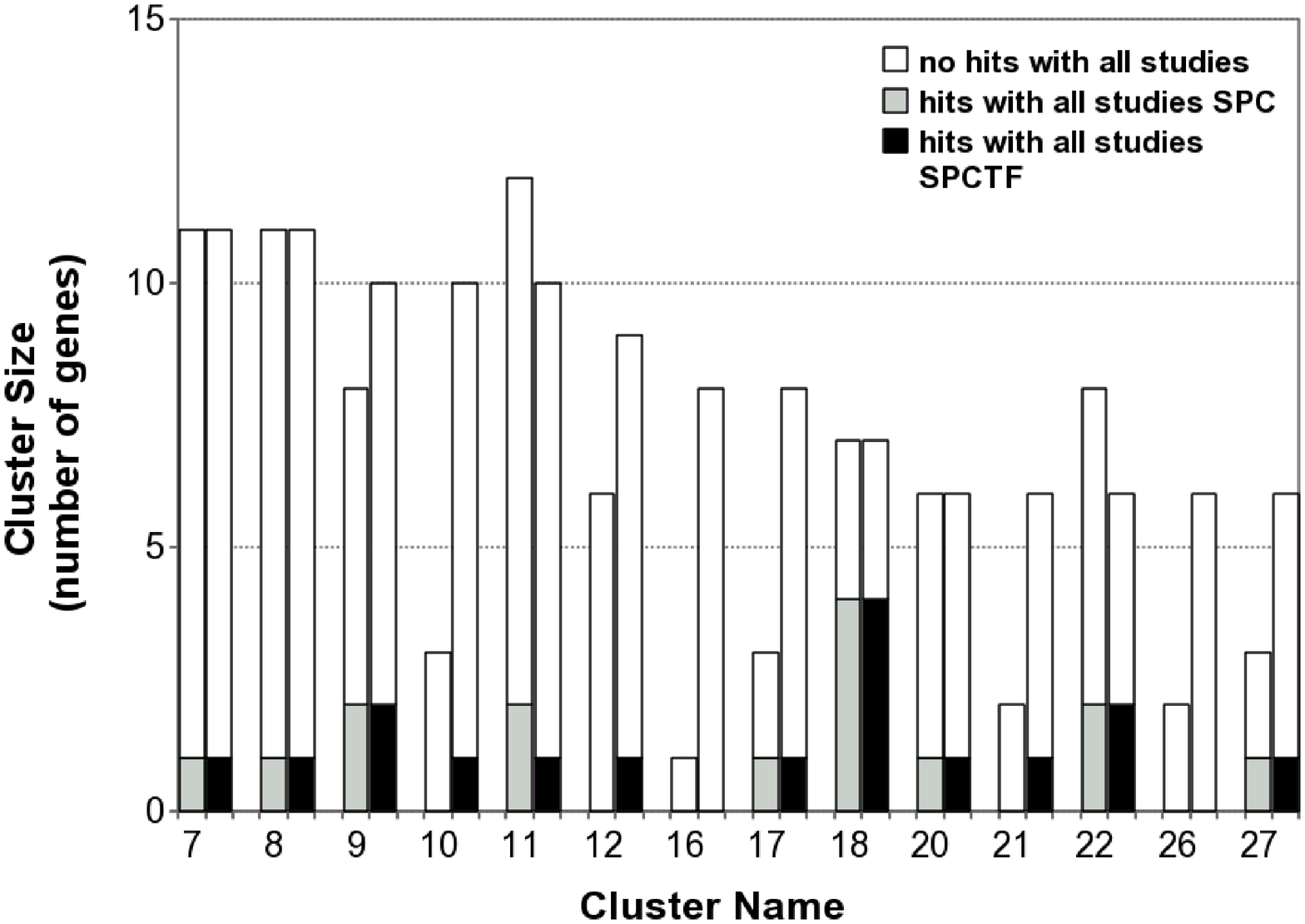}
\end{minipage}
\caption{M and N clusters, left and right respectively. Gray bars correspond to the clusters obtained with the SPC algorithm and black bars to the equivalent clusters in SPCTF.}
\label{fig:m_n_all}
\end{figure}

It is worth mentioning that more cell cycle experiments have been done since Spellman {\it et al.} \cite{Spellman} and new genes have been classified meanwhile as cell cycle regulated. Some of these newly reported cell cycle genes were obtained by Cho {\it et al.} \cite{Cho}, Pramila {\it et al.} \cite{Pramila}, Rowicka {\it et al.} \cite{Rowicka} and Lichtenberg {\it et al.} \cite{Lichtenberg}. We analize our $27$ clusters taking now as hits, genes reported either by Spellman {\it et al.} \cite{Spellman} or by one of the above mentioned studies. In this way, we gained thirty additional hits in the SPC clusters, while in SPCTF clusters we have fifty-two extra genes. The results including all the aforementioned cell cycle studies are presented in Figs. \ref{fig:all_studies1}--\ref{fig:m_n_all} \cite{list}.

\begin{figure}[!htp]
\begin{minipage}[t]{9.0cm}
\begin{center}
\flushleft{CC: cluster 4}\\
\includegraphics[height=56mm]{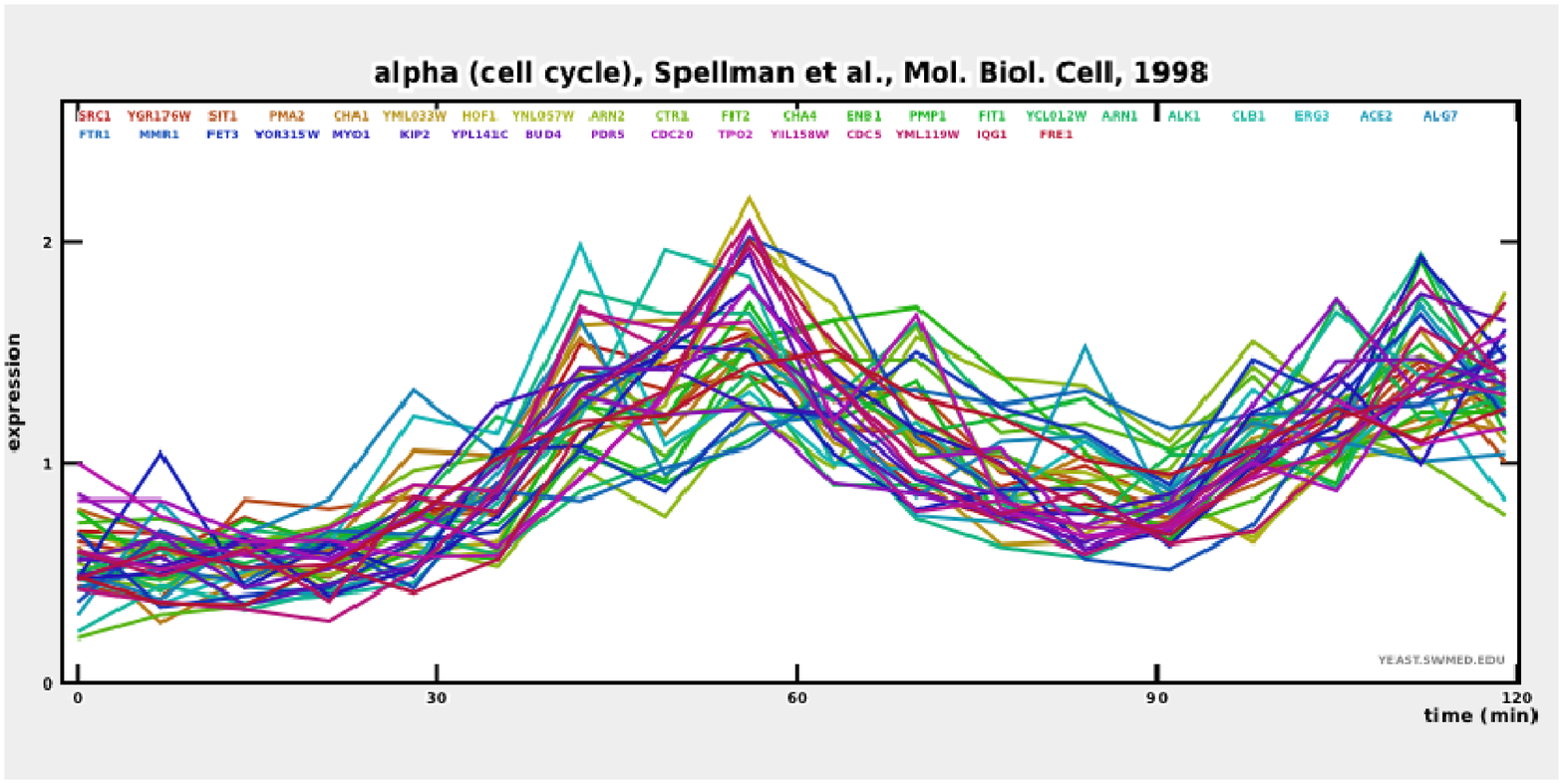}
\end{center}
\end{minipage}
\hfill
\begin{minipage}[t]{9.0cm}
\begin{center}
\flushleft{M: cluster 1}\\
\includegraphics[height=56mm]{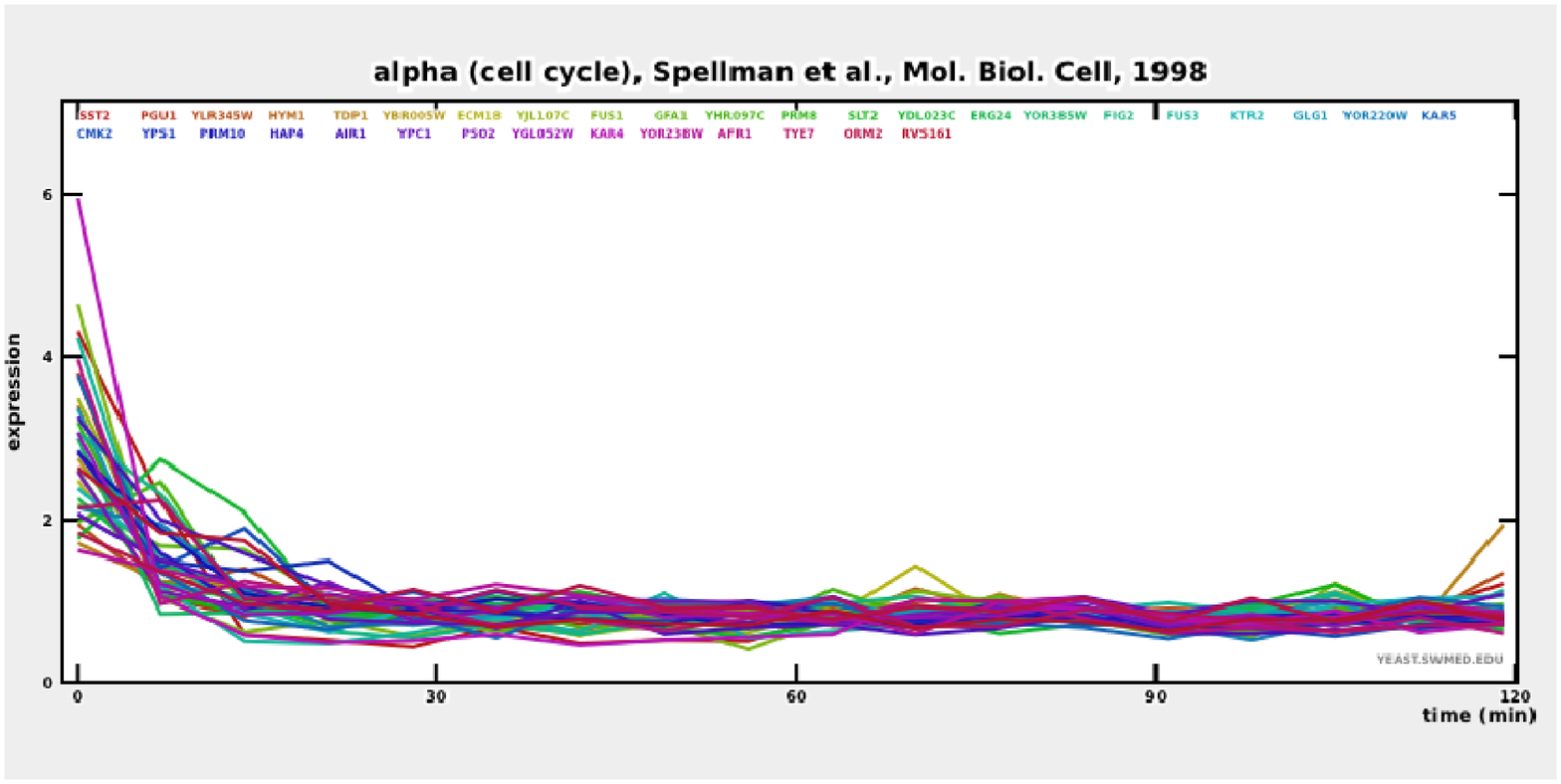}
\end{center}
\end{minipage}
\hfill
\begin{minipage}[t]{9.0cm}
\begin{center}
\flushleft{N: cluster 16}\\
\includegraphics[height=56mm]{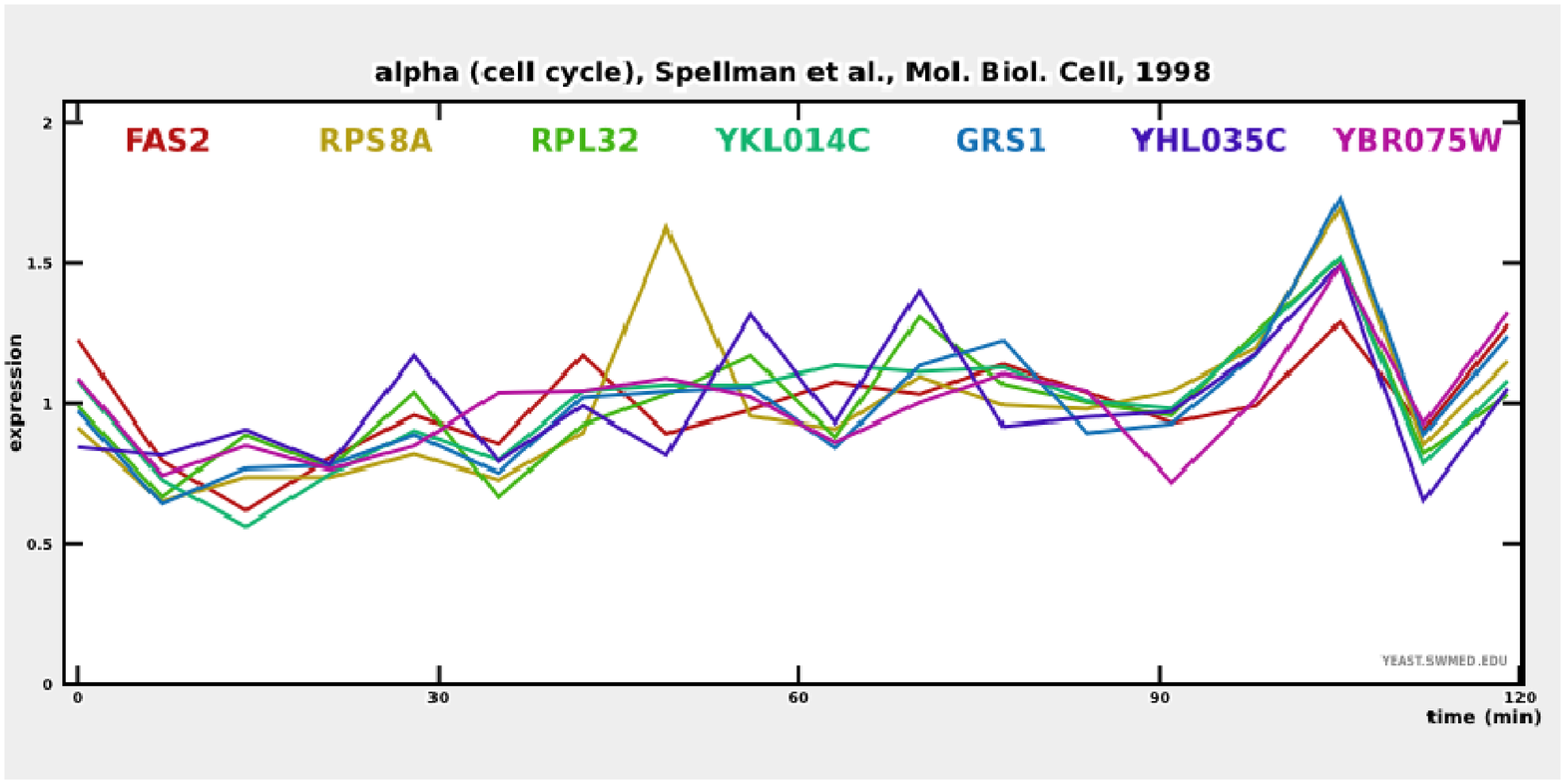}
\end{center}
\end{minipage}
\caption{(Color online) Expression profiles for a representative member of each cluster type using the SCEPTRANS tool. Expression profiles for all clusters are available in the supplementary information.}
\label{fig:exp_pro}
\end{figure}

In addition, we analyze the expression profiles of the genes conforming each cluster using the SCEPTRANS tool \cite{sceptrans}, and we notice that all the genes grouped in the same cluster had the same expression pattern. This gives us further confidence that our algorithm is grouping data correctly. The expression profiles for a representative member of each cluster type are shown in Fig. \ref{fig:exp_pro}. We also find two clusters ($21$ and $27$) that present an oscillating behaviour that is due to an artifact in the manner the microarray experiment was performed, see \cite{Ahnert, Conlon}. In the supplementary information file, we include the list of oscillating genes identified in \cite{Ahnert} and the number of these genes inside each of our first $27$ clusters. We also include the expression profiles of these clusters as well as those of size $5$ and $4$ which contain hits with cell cycle genes identified by Spellman {\it et al.} \cite{Spellman}. These clusters have also similar expression profiles but were not further analyzed because of their low number of elements. In the case of gene annotation, it is important to have clusters of many elements to effectively assure that an unknown gene shares the biological function already assigned to the other genes in the same cluster.

The CC clusters are almost entirely composed of cell cycle regulated genes reported either by Spellman {\it et al.} \cite{Spellman} or by other authors, besides, their expression patterns are similar, which leaves no doubt on their validity. For the M and N clusters, we know that they are well grouped because their elements share the same expression patterns, but in order to select those of worth for further analysis (for example in a laboratory experiment) we analyze them through MUSA, motif finding using an unsupervised approach algorithm, that can be found at {\it www.yeastract.com}. This program searches for the most common sequences (motifs) in the regulatory region of a set of genes, and compare them to the transcription factor binding sites already described in yeastract database \cite{MUSA2, MUSA3}. Results of this analysis are shown in Table \ref{tabla2}, which includes the quorum or percentage of genes containing a motif in each cluster, and the alignment score, which quantifies the level of similarity between the encountered motif and the known transcription factor associated with it. The clusters that probably would give us the best results would be those associated with cell cycle transcription factors with high percentages and scores. We select in this way, the clusters $1$, $5$, $9$, $12$, $16$ and $24$ because they have percentages higher than $70\%$ and scores higher than $80\%$.

In order to validate the MUSA analysis, we also constructed various clusters with sizes ranging from six to thirty-seven genes that were composed by genes selected at random from the original data. When analyzing these random clusters in the same way in MUSA, we obtain at most two cell cycle transcription factor coincidences.

\begin{table}[h]
{\bf MUSA analysis}
\begin{center}
\resizebox{0.85\textwidth}{!}{
\begin{tabular}{ccll}
\hline
\hline
 Cluster  & Cluster & Transcription Factor Association with Promoter & Percentage of cluster genes  \\
   Name   & Type    & (alignment score/maximum possible score)     & sharing a motif, {\it i.e.} quorum   \\
\hline
1 & M & Cup2p, Mig3p, Mig2p, Mig1p, Arg80p (5/6) & 91.67 $\%$  \\ \cline{3-4}
  &   & {\bf Swi4p} (6/6), Azf1p, {\bf Ime1p}, Dal82p, Dal81p (5/6)&  88.89 $\%$ \\ \cline{3-4}
  &   &  {\bf Ste12p} (7/7), Rox1p (6/7)& 83.33 $\%$\\
\hline
5 & M & Hac1p (6/6), {\bf Rme1p}, Arg80p, Mot3p (5/6) & 76.47 $\%$ \\
\hline
7 & N & Azf1p, Zap1p (6/7) & 81.82 $\%$\\
\hline
8 & N & Low scores & Low percentages\\
\hline
9 & N & Mig3p, Mig1p, Crz1p, Mig2p (5/6) & 80 $\%$ \\ \cline{3-4}
  &   & {\bf Rfx1p}, Arg81p (6/7) & 70 $\%$ \\
\hline
10 & N & Low scores & \\
\hline
11 & N & Azf1p (6/7) & \\
\hline
12 & N & Azf1p (7/8) &  88.89 $\%$ \\ \cline{3-4}
   &   & {\bf Rfx1p}, Cup2p (5/6) & 77.78 $\%$ \\
\hline
14 & M & Azf1p (7/8) & 75 $\%$ \\
\hline
15 & M & Low scores & Low percentages \\
\hline
16 & N & {\bf Mcm1p} (5.25/6), Crz1p (5/6) & 100 $\%$  \\ \cline{3-4}
   &   & Hap1p (5/6) & 71.43 $\%$ \\
\hline
17 & N & Arg81p, Upc2p, Sip4p, Rox1p, Crz1p, Zap1p (5/6)& 100 \\ \cline{3-4}
   &   & Pdr8p (5.33/6) & 87.5 $\%$ \\
\hline
18 & N & Azf1p, Zap1p (6/7) & 100 $\%$\\
\hline
20 & N & Low scores & \\
\hline
21 & N & Low scores & \\
\hline
22 & N & & Low percentages \\
\hline
23 & M & Low scores & \\
\hline
24 & M & Hap1p (6/6), Ecm22p, Upc2p (5/6) & 100 $\%$ \\ \cline{3-4}
   &   & {\bf Rfx1p} (6/7) & 83.33 $\%$ \\
\hline
25 & M & Hac1p (6/7) &  83.33 $\%$ \\
\hline
26 & N & Dal80p, Gat1p, Gln3p, Gzf3p (6/7)& 83.33 $\%$ \\
\hline
27 & N & Ino4p (6.5/7), Ino2p (6/7) & 100 $\%$ \\
\hline
\hline
\end{tabular} }
\end{center}
\caption{Results for quorum higher than $70\%$ and scores higher than $80\%$. Transcription factors associated to cell cycle are shown in bold. The most confident clusters are taken as those that included cell cycle transcription factor.}
\label{tabla2}
\end{table}

\section{Summary and conclusions}

Large amounts of biological information are constantly obtained by throughput techniques and clustering algorithms have taken an important place in the unraveling of this information. However, the clustering analyses offer a difficult challenge because any data set can be grouped in numerous ways, depending on the level of resolution asked for and the applied similarity measure. In this work, we propose the use of available biological information in order to strengthen the interaction between genes which share a transcription factor involved in any metabolic process, improving the similarity measure. This information is introduced in the natural evolution of the SPC algorithm, and in this way, we are able to enhance the creation and endurance of groups of possible coregulated genes. As the network spanned by the transcription factors information connects all genes, clustering directly {\it a posteriori} using only this information in the present case results into a single massive cluster (See section IV of the Supplementary Information). However, by having the distance play an important weight in the interaction formula, the far-located clusters will not join, despite sharing transcription factors between their genes.

With this in mind, we have modified the SPC algorithm, and applied both the original and modified SPCTF algorithm to one of the three Spellman {\it et al.} \cite{Spellman} data sets of the yeast cell cycle. The expression profiles of the genes in all resulting clusters show a similar behavior, but we obtain larger clusters with SPCTF. We classified them in three types, CC, M, and N, depending on the amount of cell cycle reported elements inside each cluster. With SPCTF, the CC type clusters increase in size including more cell cycle genes, and for the M and N type clusters, we also looked for common sequences in its regulatory regions and selected various groups worth of further research in order to report possible new cell cycle genes. As expected, some of these clusters include already known cell cycle genes sharing a transcription factor, {\it but more importantly, at the predictive level, they promote the inclusion of new genes with similar expression patterns}. It is also important to note that the modified algorithm can be applied to any data set, and the followed methodology leads to the selection of the potential gene subsets feasible to be experimentally investigated. Our work can serve as an example of how the inclusion of available biological information, such as transcription factors, and bioinformatic tools, such as MUSA, can lead to better and more confident results, aiding in the analysis of data coming from microarray experiments.

\section*{Acknowledgments}
\noindent The authors thank Drs. S. Ahnert and G. Sherlock for useful discussions and comments.
We also thank CONACYT for providing support for two of the authors
(M.P.M.A. and J.C.N.M.) and also the referees for helpful remarks and information. This work was partly supported through the project SEP-CONACYT-2005-49039.


\begin{thebibliography}{27}
\bibitem{list} The list of the twenty-seven clusters with the names of all their genes and identified hits is available on request from the authors.
\end{thebibliography}
\end{document}